\documentclass[conference]{IEEEtran}
\IEEEoverridecommandlockouts
\usepackage{cite}
\usepackage{amsmath,amssymb,amsfonts}
\usepackage{algorithmic}
\usepackage{graphicx}
\usepackage{textcomp}
\usepackage{xcolor}
\usepackage{booktabs}
\usepackage{float}
\usepackage{tcolorbox}
\usepackage{multirow}

\def\BibTeX{{\rm B\kern-.05em{\sc i\kern-.025em b}\kern-.08em
    T\kern-.1667em\lower.7ex\hbox{E}\kern-.125emX}}
\begin{document}

% \title{Assessing LLMs in the context of Data Science coding problems: An Empirical Investigation}

%\title{Comparative Evaluation of Large Language Models on Data Science Coding Problems: A Controlled Experiment }

%\title{Benchmarking Large Language Models on \\Data Science Coding: A Controlled Experiment}

%\title{Benchmarking LLMs for Data Science Code Generation: A Controlled Experiment}

%\title{Investigating the Efficacy of LLM-Generated Code for Data Science: A Controlled Experiment}

%\title{LLM-Generated Code for Data Science: \\A Controlled Investigation of Efficacy}

%\title{LLM-Generated Code for Data Science: A Controlled Experiment}

%\title{LLM-Generated Code for Data Science: A Comparative Study}

%\title{Evaluating LLM-Generated Code for Data Science: A Controlled Experiment}
%Findings
%\title{LLM-Generated Code for Data Science: \\Insights from a Comparative Controlled Experiment}
%\title{How Effective are LLMs for Data Science Coding?  A Controlled Experiment}  ***SUBMITTED TITLE
\title{LLM4DS: Evaluating Large Language Models for Data Science Code Generation}

%\title{LLM4DS: Empirical Evaluation of Large Language Models for Data Science Code Generation}
% Future work for paper extension for Journal: Benchmarking Large Language Models for Data Science Code Generation - 
%Benchmarking LLMs through an Empirical Evaluation 

%\title{LLM4DS: An Empirical Evaluation of Large Language Models for \\Data Science Code Generation}
%\title{LLMs in Data Science: \\A Controlled Experiment on Code Generation}
%\title{LLMs for Data Science Coding: A Comparative Controlled Study}

\author{
    \IEEEauthorblockN{Nathalia Nascimento}
    \IEEEauthorblockA{\textit{EASER, Eng. Division} \\
    \textit{Pennsylvania State University}\\
    Great Valley, USA \\
    nqm5742@psu.edu}
    \and 
    \IEEEauthorblockN{Everton Guimaraes}
    \IEEEauthorblockA{\textit{EASER, Eng. Division} \\
    \textit{ Pennsylvania State University}\\
    Great Valley, USA \\
    ezt157@psu.edu}
    \and
    \IEEEauthorblockN{Sai Sanjna Chintakunta}
    \IEEEauthorblockA{\textit{EASER, Eng. Division} \\
    \textit{ Pennsylvania State University}\\
    Great Valley, USA \\
    sqc6557@psu.edu}
    \and
    \IEEEauthorblockN{Santhosh Anitha Boominathan}
    \IEEEauthorblockA{\textit{EASER, Eng. Division} \\
    \textit{Pennsylvania State University}\\
    Great Valley, USA \\
    sfa5971@psu.edu}
    % \IEEEauthorblockN{Anonymous Submission}
    % \IEEEauthorblockA{\textit{xxxx} \\}
}

\maketitle

\begin{abstract}
The adoption of Large Language Models (LLMs) for code generation in data science offers substantial potential for enhancing tasks such as data manipulation, statistical analysis, and visualization.
However, the effectiveness of these models in the data science domain remains underexplored. 
This paper presents a controlled experiment that empirically assesses the performance of four leading LLM-based AI assistants—Microsoft Copilot (GPT-4 Turbo), ChatGPT (o1-preview), Claude (3.5 Sonnet), and Perplexity Labs (Llama-3.1-70b-instruct)—on a diverse set of data science coding challenges sourced from the Stratacratch platform. Using the Goal-Question-Metric (GQM) approach, we evaluated each model’s effectiveness across task types (Analytical, Algorithm, Visualization) and varying difficulty levels. 
Our findings reveal that all models exceeded a 50\% baseline success rate, confirming their capability beyond random chance. Notably, only ChatGPT and Claude achieved success rates significantly above a 60\% baseline, though none of the models reached a 70\% threshold, indicating limitations in higher standards. ChatGPT demonstrated consistent performance across varying difficulty levels, while Claude's success rate fluctuated with task complexity. 
Hypothesis testing indicates that task type does not significantly impact success rate overall. For analytical tasks, efficiency analysis shows no significant differences in execution times, though ChatGPT tended to be slower and less predictable despite high success rates. For visualization tasks, while similarity quality among LLMs is comparable, ChatGPT consistently delivered the most accurate outputs.
This study provides a structured, empirical evaluation of LLMs in data science, delivering insights that support informed model selection tailored to specific task demands. Our findings establish a framework for future AI assessments, emphasizing the value of rigorous evaluation beyond basic accuracy measures. 

\end{abstract}


\section{Introduction}

Large Language Models (LLMs) have emerged as transformative tools with the potential to revolutionize code generation in various domains, including data science \cite{halevy2023will, nascimento2023gpt, li2024can, lai2023ds, kazemitabaar2024improving}. Their ability to generate human-like text and code opens up possibilities for automating complex tasks in data manipulation, visualization, and analytics. As data science projects often require extensive coding efforts that are time-consuming and demand significant expertise, leveraging LLMs could greatly enhance productivity and accessibility in this field. However, the effectiveness and reliability of LLM-generated code for data science applications remain underexplored, necessitating a thorough evaluation.

While previous studies have evaluated LLMs in general programming tasks using platforms like LeetCode \cite{nguyen2022empirical, nathalia2023artificial, kuhail2024will, coignion2024performance}, the HumanEval benchmark \cite{chen2021evaluating}, and GitHub Projects \cite{grewal2024analyzing}, Gu et al. \cite{gu2024effectiveness} identified a notable gap in approaches to evaluate domain-specific code generation. They demonstrated that LLMs exhibit sub-optimal performance in generating domain-specific code for areas such as web and game development, due to their limited proficiency in utilizing domain-specific libraries. This finding underscores the need for more focused evaluations that consider the unique challenges of specialized domains like data science, which involve tasks such as handling datasets, performing complex statistical analyses, and generating insightful visualizations—areas not fully represented in general programming assessments.

This paper addresses this gap by providing an empirical evaluation \cite{wohlin2012experimentation} of multiple LLMs on diverse data science-specific coding problems sourced from the Stratascratch platform \cite{stratascratch_master_coding}. The controlled experiment involves four main steps: (i) selecting 100 Python coding problems from Stratascratch, distributed across three difficulty levels (easy, medium, hard) and three problem types (Analytical, Algorithm, Visualization); (ii) transforming these problems into prompts following the optimal prompt structure for each type; (iii) using these prompts for each AI assistant to generate code solutions; and (iv) evaluating the generated code based on correctness, efficiency, and other relevant metrics.

Our research seeks to answer the following question: How effective are LLMs for data science coding? By systematically assessing the performance of these AI assistants, we aim to identify their strengths and limitations in automating code generation for data science problems.

\textbf{Our contributions are multifold:} \begin{enumerate} \item We provide an empirical evaluation of multiple LLMs on data science-specific coding problems, filling a critical gap in current research. \item We assess Stratacratch as a platform to benchmark LLMs for data science code generation, evaluating its suitability and potential as a standardized dataset for LLM performance in this domain. \item We analyze the success rate of these models across different task categories—Analytical, Algorithm, and Visualization—and difficulty levels, offering insights into their practical utility in data science workflows. \item We highlight the challenges and limitations of LLMs in this domain, providing a foundation for future improvements and research in AI-assisted data science. \end{enumerate}

This paper is organized as follows. Section 2 presents the related work. Section 3 describes the controlled experiment, outlining the research questions, hypotheses, and methodology. Section 4-5 presents the experimental results and discusses threats to validity. Sections 6-8 brings final remarks and suggestions for future work.

\section{Related Work}
%Chopra et al. \cite{chopra2023conversational} examined the use of LLMs in data science tasks like data preprocessing and analytics. They identified challenges in interacting with LLM-powered chatbots, particularly in contextual data retrieval, prompt formulation for complex tasks, and adapting generated code. To address these issues, they proposed design improvements such as data brushing and inquisitive feedback loops.

In the realm of code generation, prior studies have evaluated LLMs like ChatGPT and GitHub Copilot using platforms such as HumanEval Benchmark, LeetCode, and Github. Nascimento et al. \cite{nathalia2023artificial} compared code generated by ChatGPT against human-written solutions, assessing performance and memory efficiency. Kuhail et al. \cite{kuhail2024will} evaluated ChatGPT on 180 LeetCode problems, providing insights into its capabilities and limitations. Coignion et al. \cite{coignion2024performance} investigated different LLMs on general coding problems from LeetCode, focusing on performance metrics. Nguyen and Nadi \cite{nguyen2022empirical} assessed GitHub Copilot's code generation on 33 LeetCode problems, evaluating correctness and understandability.

Beyond traditional programming tasks, LLMs have been applied in data science-specific domains, where recent research has explored the models' capacity to handle complex queries and data manipulation tasks. Troy et al. \cite{troy2023enabling} demonstrated that LLMs could generate SQL statements for cybersecurity applications, specifically highlighting their capability in structured query generation. In another study, Malekpour et al. \cite{malekpour2024towards} introduced an LLM routing framework designed for text-to-SQL tasks, optimizing the selection of models based on cost-efficiency and accuracy. Li et al. \cite{li2024can} identified limitations even in advanced models like GPT-4, noting that these models achieved only 54.89\% execution accuracy on complex text-to-SQL queries—significantly below the human benchmark of 92.96\%. Additionally, Kazemitabaar et al. \cite{kazemitabaar2024improving} delved into the challenges of data analysis with conversational AI tools like ChatGPT, identifying difficulties users face in verifying and guiding AI-generated results for desired outcomes.

Lai et al. \cite{lai2023ds} proposed the DS-1000 benchmark, a dataset specifically crafted for evaluating code generation in data science contexts. DS-1000 comprises 451 unique data science problems sourced from StackOverflow and spans seven essential Python libraries, including Numpy and Pandas. A key feature of this benchmark is its emphasis on problem perturbations, aimed at reducing the risk of model memorization. The dataset accounts for the unique challenges of data science tasks, which often lack executable contexts, may depend on external libraries, and can have multiple correct solutions. Lai et al. demonstrated the effect of different types of problem perturbations by testing models like Codex, InCoder, and CodeGen, with the best accuracy being 43.3\% achieved by Codex-002. However, while DS-1000 provides a robust dataset for testing, Lai et al. do not perform a comparative empirical evaluation across multiple LLMs, leaving open questions about how current models fare on this benchmark.

Despite these advancements, much of the current research has been limited to either general coding tasks or SQL-specific applications. The nuances of data science problems—ranging from data manipulation and complex analyses to visualization—remain underexplored in LLM evaluations. Our work addresses this gap by conducting an empirical experiment using four leading LLMs on a set of data science problems extracted from the Stratacratch dataset, encompassing various difficulty levels and problem types. Unlike prior studies, which primarily introduce benchmarks or focus on specific task categories, our approach offers a detailed examination of LLM performance across a broader spectrum of data science challenges. 
%Through this investigation, we aim to offer a more comprehensive understanding of the strengths and limitations of LLMs in handling the diverse demands of data science coding tasks.

% Through this investigation, we aim to offer a more comprehensive understanding of the strengths and limitations of LLMs in handling the diverse demands of data science coding tasks.

%Despite the expanding role of LLMs in code generation and data science tasks, prior studies have mainly focused on general programming problems or specific applications like SQL generation, leaving a notable gap in evaluating their effectiveness on a broader spectrum of data science coding challenges that involve unique complexities such as data manipulation, advanced analyses, and visualization. Our work addresses this gap by providing an empirical evaluation of multiple LLMs on diverse data science-specific coding problems. By assessing their performance across different task categories and difficulty levels, we contribute to a deeper understanding of LLM capabilities and limitations in the context of data science code generation.

\section{Controlled Experiment} \label{sec:controlled}
In line with the controlled experiment methodology by Wohlin et al. \cite{wohlin2012experimentation}, our study aims to evaluate and compare the effectiveness of four prominent LLM-based AI assistants—Microsoft Copilot (GPT-4 Turbo), ChatGPT (o1-preview), Claude (3.5 Sonnet), and Perplexity Lab (Llama-3.1-70b-instruct)—in solving data science coding tasks sourced from the Stratascratch platform \cite{stratascratch_master_coding}. 

\textbf{Effectiveness} in this context refers to the degree to which these models achieve desired outcomes across four key aspects: \textit{success rate}, \textit{efficiency}, \textit{quality of output}, and \textit{consistency}. Specifically, we define:
\begin{itemize}
    \item \textbf{Success Rate} as the proportion of correctly generated code solutions, measured by the percentage of solutions that achieve the correct result regardless of the number of attempts;
    \item \textbf{Efficiency} as the runtime execution speed of the generated solution;
    \item \textbf{Quality of Output} as the alignment of generated solutions with expected outcomes, particularly for visualization tasks;
    \item \textbf{Consistency} as the reliability of each model's performance across varying difficulty levels and task types.
\end{itemize}

%Our primary objective is to determine how effectively each LLM performs in generating accurate, efficient, and high-quality solutions, thereby highlighting potential strengths and weaknesses among these models. 

\subsection{Research Questions, Hypotheses, and Metrics}
To systematically explore effectiveness, we structured our investigation around specific research questions, each accompanied by testable hypotheses and relevant evaluation metrics. Table \ref{tab:research_questions_hypotheses_metrics} details these research questions, hypotheses, and corresponding metrics.

\begin{table*}[!ht]
%\small % Reduce font size within the table
\centering
\caption{Research Questions, Hypotheses, and Metrics}
\begin{tabular}{|p{5.2cm}|p{3.7cm}|p{3.7cm}|p{3.7cm}|}
\hline
\textbf{Research Question} & \textbf{Null Hypothesis} & \textbf{Alternative Hypothesis} & \textbf{Metrics} \\
\hline
RQ1: How successful are LLMs in solving data science coding problems, and do they outperform each other in success rate? & 
H0\_1: The success rate of each LLM in solving data science coding problems is not significantly higher than random chance (50\%). \newline 
H0\_1a: There is no significant difference in success rates between LLM pairs. & 
H1\_1: The success rate of each LLM in solving data science coding problems is significantly higher than random chance (50\%). \newline
H1\_1a: At least one pair of LLMs shows a significant difference in success rates. & 
Overall success rate (percentage of correct solutions) and pairwise success rate comparisons \\
\hline
RQ2: Does the difficulty level of coding problems (easy, medium, hard) influence the success rate of the different LLMs, and do specific LLMs outperform others at each difficulty level? & 
H0\_2: Difficulty level does not significantly affect the success rate of the LLMs. \newline
H0\_2a: There is no significant difference in success rates between LLM pairs within each difficulty level. & 
H1\_2: The success rate of the LLMs varies significantly with difficulty level. \newline
H1\_2a: At least one pair of LLMs shows a significant difference in success rate within a specific difficulty level. & 
Success rate (percentage of correct solutions) across difficulty levels and pairwise success rate comparisons within each level \\
\hline
RQ3: Does the type of data science task (Analytical, Algorithm, Visualization) influence the success rate of the different LLMs, and do specific LLMs outperform others for certain task types? & 
H0\_3: The type of data science task does not significantly impact the LLMs' success rate. \newline
H0\_3a: There is no significant difference in success rates between LLM pairs within each task type. & 
H1\_3: The success rate of the LLMs varies significantly with the type of data science task. \newline
H1\_3a: At least one pair of LLMs shows a significant difference in success rate within a specific task type. & 
Success rate (percentage of correct solutions) for each task type and pairwise success rate comparisons within each type \\
\hline
RQ4: For Analytical questions, do the LLMs differ in the efficiency (execution time) of the code they generate? & 
H0\_4: The population medians of the execution times across the LLMs for Analytical questions are equal. & 
H1\_4: At least one LLM has a different population median execution time for Analytical questions compared to others. & 
Execution time for each generated solution on Analytical problems, per LLM \\
\hline
RQ5: For visualization tasks, do the LLMs differ in the quality (similarity) of the visual outputs they produce compared to expected results? & 
H0\_5: The population medians of the similarity scores for visualization outputs across the LLMs are equal. & 
H1\_5: At least one LLM has a different population median similarity score for visualization outputs compared to others. & 
Similarity scores for visualization outputs compared to expected results \\
\hline
\end{tabular}
\label{tab:research_questions_hypotheses_metrics}
\end{table*}

\subsection{Variables Selection}

To structure our analysis, we identified key variables that allow us to examine the performance of each AI assistant across different problem types and difficulty levels. 

The \textbf{independent variables} in this study, which we controlled or varied, include:
\begin{itemize}
    \item \textbf{LLM-based AI assistants}: The four AI models under evaluation—Microsoft Copilot, ChatGPT, Claude, and Perplexity Lab.
    \item \textbf{Difficulty level of coding problems}: Easy, Medium, Hard.
    \item \textbf{Type of Data Science task}: Analytical, Algorithm, Visualization.
\end{itemize}

The \textbf{dependent variables} are the metrics we measured to assess each AI assistant's effectiveness:
\begin{itemize}
    \item \textbf{Success rate}: The percentage of correct solutions generated by each LLM, regardless of the number of attempts.
    \item \textbf{Running time}: Execution time of code for Analytical questions.
    \item \textbf{Graph similarity scores}: Similarity between generated and expected graphs for Visualization questions.
\end{itemize}

These variables connect directly to the research questions and metrics outlined in Table \ref{tab:research_questions_hypotheses_metrics}, allowing us to systematically investigate the impact of each independent variable on the AI models' performance. %By measuring these dependent variables, we can assess how effectively each model handles tasks of different types and complexities.

\section{Experiment Operation}

The experiment was conducted over the span of two months. For each of the four LLMs, we generated a solution for each of the 100 selected problems, resulting in a total of 400 generated coding solutions. Two researchers manually interacted with the AI assistants by inputting the prompts into their respective interfaces. They then copied the generated code and submitted it to the Stratascratch platform to assess its correctness and functionality. The researchers recorded whether the solution worked as intended and noted any necessary adjustments. Since Stratascratch provides execution time only for Analytical questions and similarity scores for Visualization questions, we collected these specific measurements accordingly.

\begin{figure}[H]
    \centering
    \includegraphics[width=0.9\linewidth]{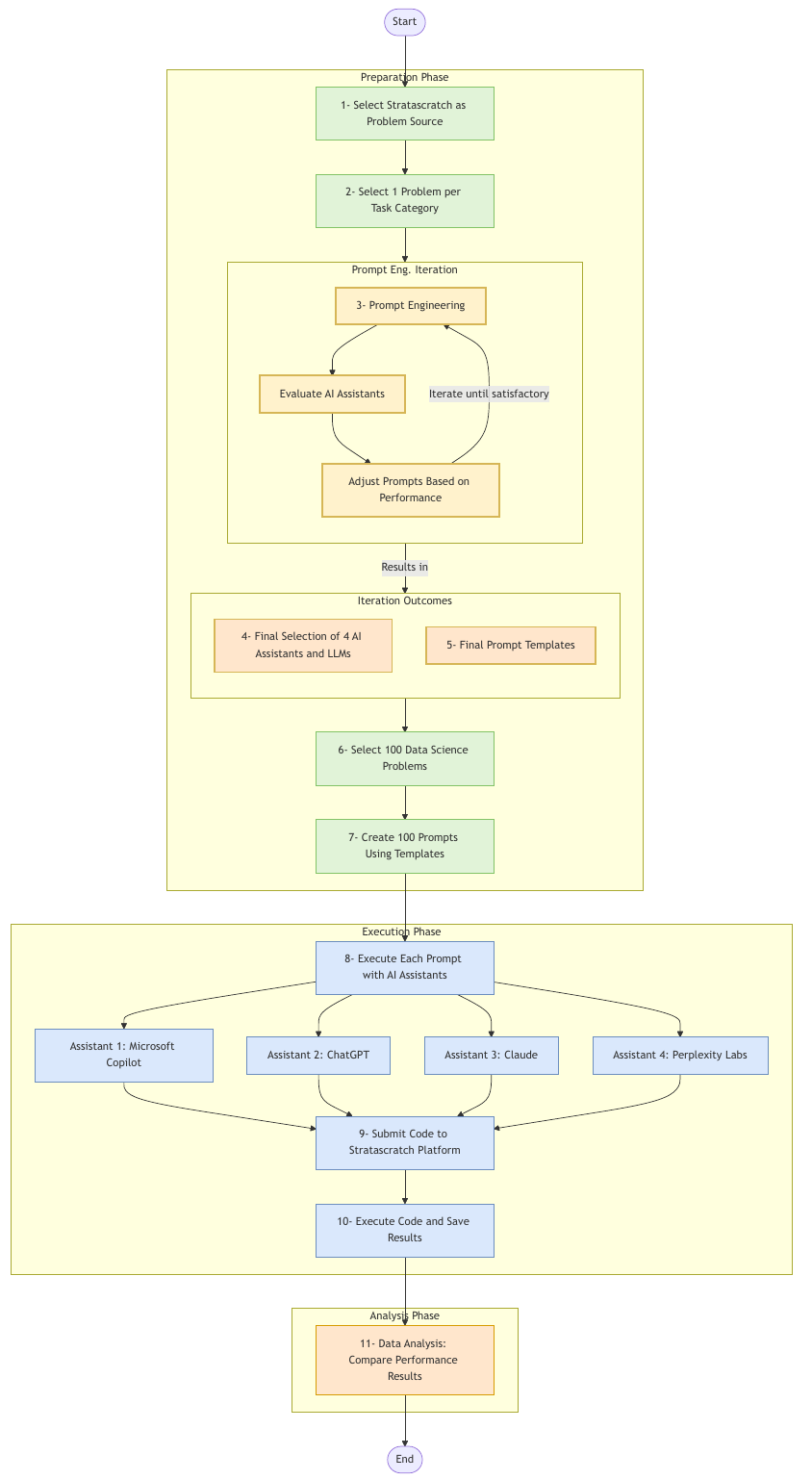}
    \caption{Overview of the Experimental Process.}
    \label{fig:overview}
\end{figure}

 The overall process of our controlled experiment consists of 11 steps, as illustrated in Figure \ref{fig:overview}:

\begin{enumerate}
    \item Select the problem source: We chose Stratascratch \cite{stratascratch_master_coding} as the platform for sourcing data science problems.
    \item Select one problem per task category for prompt engineering: One problem from each data science task category (Analytical, Algorithm, Visualization) was selected to refine our prompt templates.
    \item Prompt Engineering with feedback loop: We performed prompt engineering by iteratively adjusting the prompts and assessing the performance of different LLM versions, creating optimal prompt structures for each task type.
    \item Selection of AI assistants and LLMs: Four AI assistants, each utilizing a different LLM, were selected for the experiment.
    \item Definition of final prompts: The finalized prompt templates were established for each problem type based on the prompt engineering process.
    \item Selection of 100 Data Science problems: We selected 100 data science problems covering various topics across the three task types to ensure a comprehensive evaluation.
    \item Creation of prompts: The selected problems were incorporated into the prompt templates, resulting in 100 tailored prompts.
    \item Execution with AI assistants: Each prompt was executed using the four AI assistants, and the generated Python code was saved.
    \item Submission to Stratascratch platform: The generated code solutions were submitted to the Stratascratch platform interface for evaluation.
    \item Execution and result collection: The code was executed on Stratascratch, and the results were saved into a results dataset.
    \item Data analysis: We compared the performance results of the four LLMs to analyze their effectiveness.
\end{enumerate}

The following subsections and sections provide more detailed explanations of the main steps:
\begin{itemize}
    \item Subsection \ref{sub:dataset} describes the selection of 100 Python coding problems from Stratascratch, categorized by difficulty levels (easy, medium, hard) and types (Analytical, Algorithm, Visualization).
    \item Subsection \ref{sub:prompt} outlines the iterative prompt engineering process, including the development and refinement of prompt templates for each task type. This resulted in optimal prompt structures used to transform the selected problems into 100 tailored prompts.
    \item Subsection \ref{sec:generation} explains the process of using these prompts with each AI assistant to generate code solutions.
    \item Section \ref{sec:analysis} covers the data analysis process.
\end{itemize}

% Evaluation Metrics: We outline how the generated code was evaluated based on correctness, efficiency, and other relevant metrics.

\subsection{Dataset: Selection of Data Science Problems} \label{sub:dataset}

For our study, we selected the Stratascratch \cite{stratascratch_master_coding} platform as the source of data science coding problems. Stratascratch is a platform that aggregates real-world data science interview questions from various companies, providing a diverse set of problems that are representative of typical tasks encountered in data science, such as data manipulation, algorithm development, and data visualization. 

Stratascratch problems are organized into three difficulty levels (easy, medium, and hard) and three main types, each addressing unique aspects of data science problem-solving:

\textbf{Analytical:} These problems involve tasks requiring data analysis and manipulation using tools like \texttt{pandas} and SQL. Topics include data aggregation, filtering, conditional expressions, and data formatting. 
%Analytical problems challenge users to extract insights from data and perform complex manipulations typical in data science workflows.

\textbf{Algorithm:} These challenges focus on computational problem-solving and algorithm development. Topics in this category include array manipulation, linear regression, probability, graph theory, recursion, and optimization techniques. 
%These problems test a user's ability to design efficient algorithms, often requiring performance optimization and complexity analysis.

\textbf{Visualization:} These problems require the creation of charts and graphs to represent data insights visually. Topics cover distribution analysis, time-series trend analysis, spatial data visualization, and comparison of categorical and numerical data. 
%These tasks help assess a user's ability to effectively communicate data insights and trends through visual representation, an essential skill in data storytelling and reporting.

\begin{figure}[H] 
\centering 
\fbox{\includegraphics[width=0.95\linewidth]{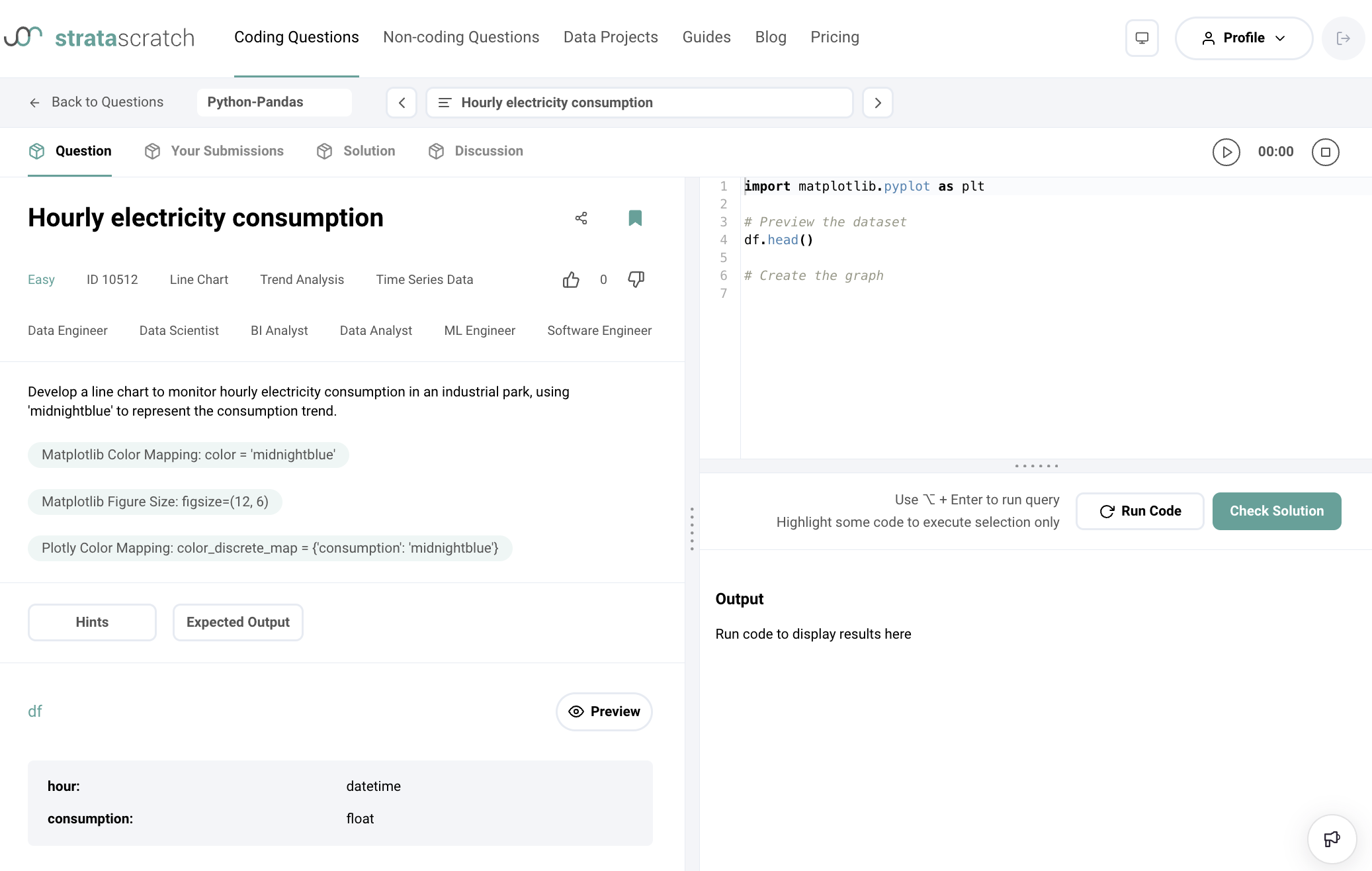}} \caption{Example of a Visualization Problem from the Stratascratch platform.} 
\label{fig:strata} 
\end{figure}

An example Stratascratch problem is shown in Figure~\ref{fig:strata}, demonstrating the typical interface and information available for each question.
 
To build our dataset, we used random sampling while ensuring balanced representation across problem types and difficulty levels—selecting 100 Python coding problems in total, with 35 Analytical, 35 Algorithm, and 30 Visualization problems. From these 100 questions, 34 are easy, 32 are medium, and 34 are hard. 
To avoid infringing any intellectual property from Stratascratch, we omitted the full problem descriptions from our dataset. However, we have provided a table containing problem IDs, difficulty levels, links, and topic descriptions, which gives sufficient context for each task (dataset available in \cite{EASER_LLM4Code_Study4_2023}).
%This resource allows for replication of our study while respecting proprietary rights.

\subsection{Prompt Engineering: Transforming Problems into Prompts} \label{sub:prompt}

\begin{tcolorbox}[title=Prompt Template for Visualization Problems, colback=white, colframe=black]
% \label{prompt}
\scriptsize
Act as a data scientist and provide working Python3 code for the problem below.

\textbf{CRITICAL REQUIREMENTS:}

\begin{enumerate}
    \item Use \textbf{ONLY} the exact prefilled code snippet as starting point.
    \item \textbf{NO} additional imports beyond what's given.
    \item \textbf{NO} sample/test data creation—use \textbf{ONLY} the provided DataFrame.
    \item \textbf{NO} functions unless explicitly required in the original code.
    \item Code must end with appropriate visualization command (\texttt{plt.show()} for Matplotlib, \texttt{fig.show()} for Plotly, etc.).
    \item Code must be fully runnable without any modifications.
\end{enumerate}

\textbf{PROBLEM:}

Title: \textbf{\{\texttt{Title}\}}

Description: \textbf{\{\texttt{Description}\}}

Difficulty: \textbf{\{\texttt{Level}\}}

\textbf{\{\texttt{DataFrame}\}}

\textbf{\{\texttt{Additional}\}}

\textbf{EXPECTED OUTPUT FORMAT:}

\begin{itemize}
    \item Direct plotting code only.
    \item Must end with appropriate \texttt{show()} command.
    \item No \texttt{return} statements.
    \item No \texttt{print} statements.
    \item No functions unless explicitly required.
    \item No test data creation.
\end{itemize}

\textbf{Prefilled code snippet (use exactly):}

\textbf{\{\texttt{Code}\}}
\end{tcolorbox}

This step started by selecting one problem from each task category-Analytical, Algorithm, and Visualization-for prompt development. These problems were outside our main dataset to avoid biasing the evaluation results. During this phase, we experimented with various prompt structures and observed the models' outputs. Initially, the models often generated code that included datasets or functions not specified in the problem descriptions. To address this, we iteratively refined the prompts by introducing specific instructions and constraints.

We tested several LLMs during prompt engineering, including some not selected for the main experiment. Some LLMs, such as Gemini (1.5 Flash), could not produce functional code even for easy problems, despite multiple prompt refinements. Others, like YouChat Pro, was capable but was not included in the final selection to avoid redundancy.

To ensure consistency and minimize subjectivity, we automated the conversion of problem descriptions into prompts. This involved creating prompt templates tailored to each problem type-Analytical, Algorithm, and Visualization-which addressed the unique requirements of each category. %Developing distinct templates standardized the input across all problems and AI assistants, ensuring that the models received clear and consistent instructions.
This section illustrates the prompt template used for Visualization problems. The templates for Algorithm and Analytics tasks are available in \cite{EASER_LLM4Code_Study4_2023}. Our automated prompt generation system parsed the problem descriptions and inserted the information into the appropriate template based on the problem type.

\subsection{Code Generation and Execution} \label{sec:generation}
In this experiment, we presented 100 problem prompts to four AI assistants—Microsoft Copilot, ChatGPT, Claude, and Perplexity Labs—generating a total of 400 code solutions (100 problems per assistant). For each problem, a new chat thread was initiated with the AI assistant to ensure no influence from previous interactions. Each AI assistant was given up to three attempts per problem, guided by feedback such as ``Not worked" (which yielded better results with ChatGPT and Copilot) or ``Wrong answer" (more effective with Claude and Perplexity) to prompt improvements.

\begin{figure}[H] 
\centering 
\fbox{\includegraphics[width=0.90\linewidth]{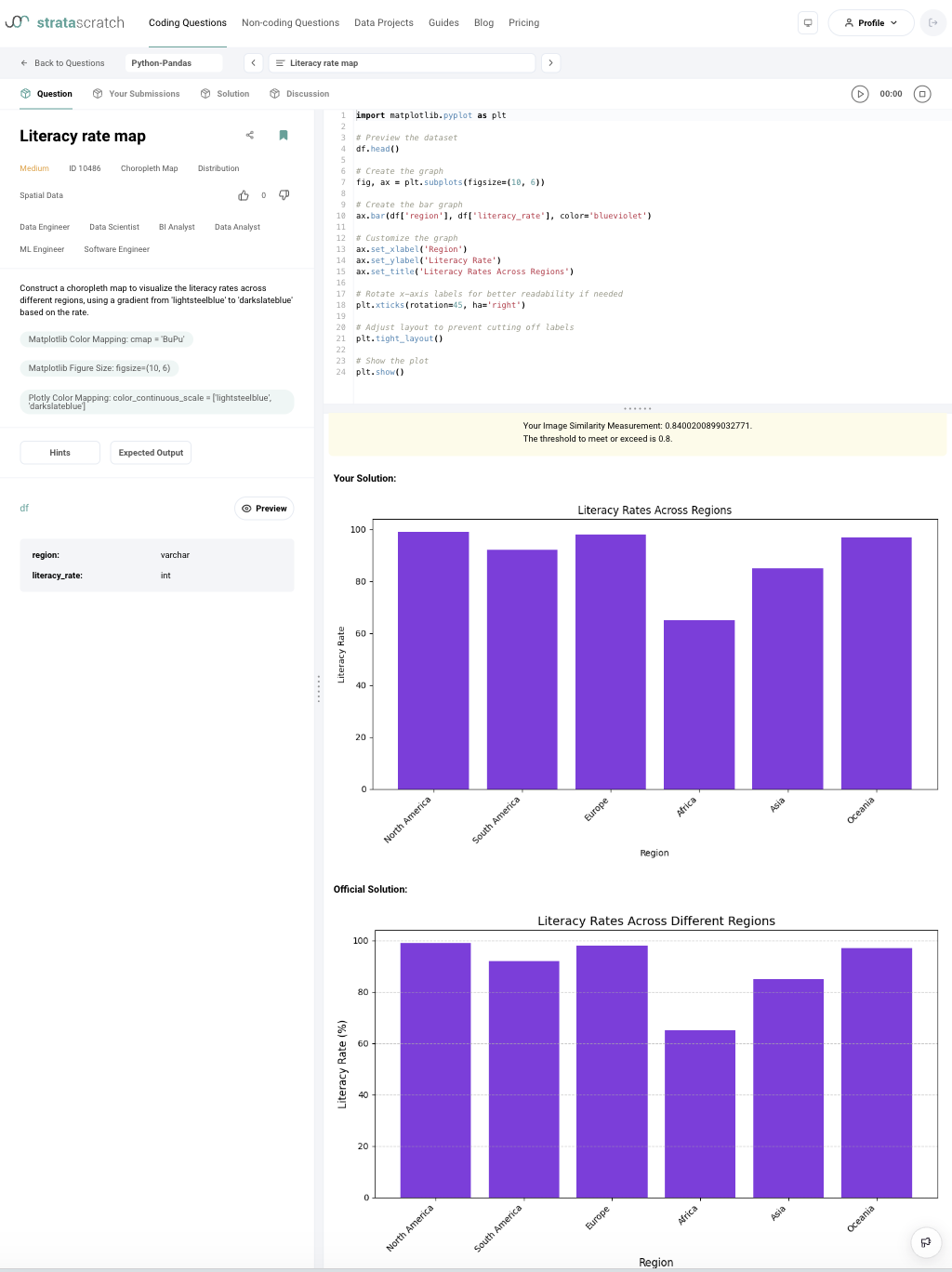}} 
\caption{Similarity comparison for a Visualization Problem.}
\label{fig:similarity-comparison} 
\end{figure}

To evaluate the solutions, we executed them on the Stratascratch platform and recorded the metrics provided by the platform, depending on the type of problem. For visualization problems, for example, the platforms calculates the similarity of the generated graphs with the expected outputs. Figure~\ref{fig:similarity-comparison} illustrates a similarity comparison between a graph generated by the Perplexity model and the expected solution provided by Stratascratch.

Due to Stratascratch platform constraints (e.g., limitations on library imports and required code formatting), we allowed minor manual edits to adapt the AI-generated code for consistent evaluation. These adjustments included removing prohibited imports (e.g., import os), modifying code structure (e.g., removing function encapsulation when global code was needed), and eliminating unnecessary print statements in favor of returns. These edits preserved the core logic and functionality of the solutions and were documented for transparency and reproducibility. This documentation (available in \cite{EASER_LLM4Code_Study4_2023}) includes the nature of the edits and their reason.

\section{Analysis and Interpretation} \label{sec:analysis}

This section presents the statistical analysis of data collected during the experiment. The dataset includes information such as problem IDs, the code generated by each LLM, and associated performance metrics, which is available in full for reproducibility in \cite{EASER_LLM4Code_Study4_2023}.

\begin{figure}[!ht]
    \centering
    \includegraphics[width=0.95\linewidth]{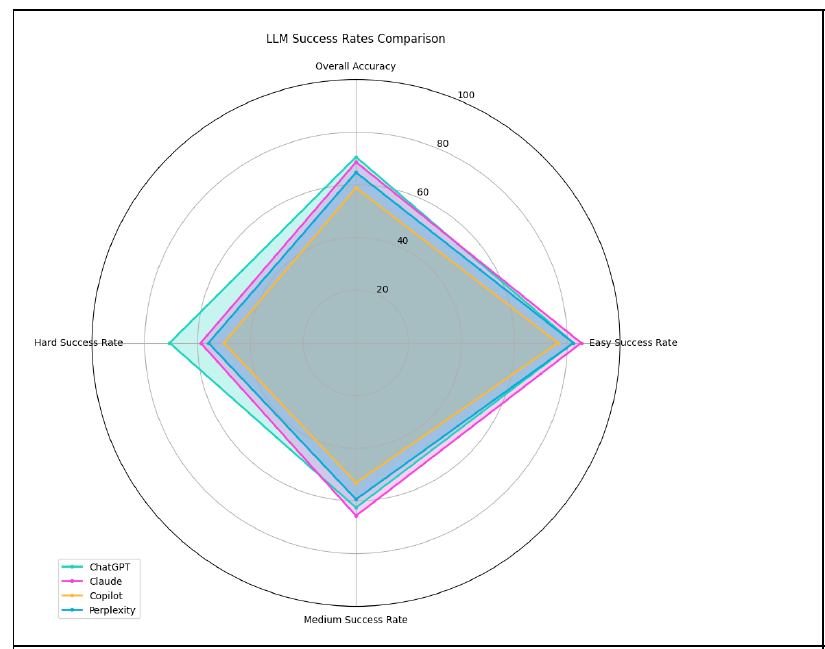}
    \caption{Overall Success Rate of LLMs.}
    \label{fig:fig1}
\end{figure}

Figure \ref{fig:fig1} provides a general overview of the LLMs' assertiveness across all tasks and difficulty levels. This initial visualization offers a preliminary look at overall trends, while more detailed analyses follow for each research question (RQ).

For each research question (RQ), we begin by visualizing the data to provide an intuitive understanding of the performance distributions across different conditions. In addition to visualization and descriptive statistics, we perform hypothesis testing for each RQ. 

\subsection{RQ1: Success Rate of LLMs in Solving Data Science Problems}

\begin{figure}[!ht]
    \centering
    \includegraphics[width=0.90\linewidth]{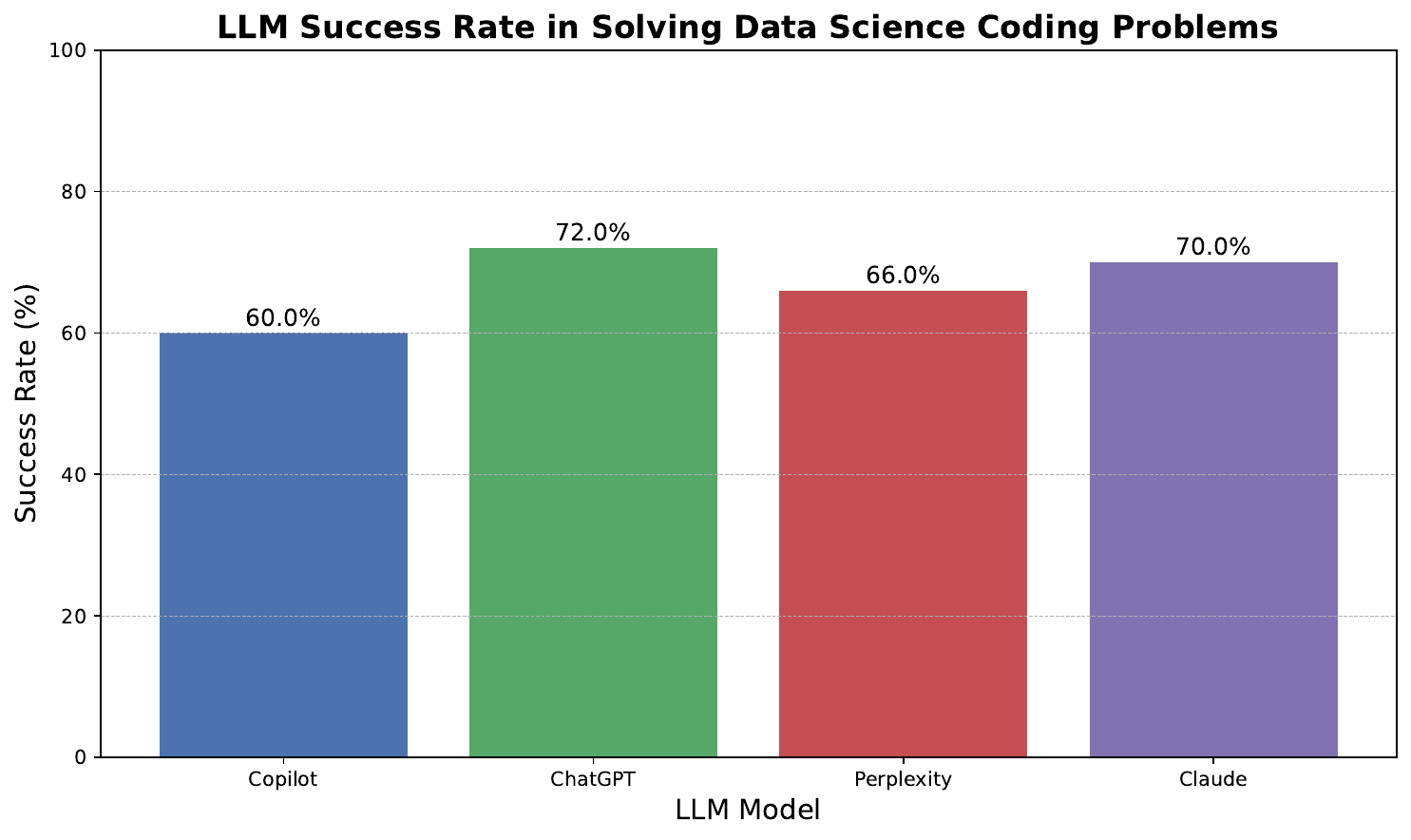}
    \caption{RQ1 - LLM success rate in solving DS coding problems.}
    \label{fig:overall-success-rate}
\end{figure}

As shown in Figure \ref{fig:overall-success-rate}, ChatGPT achieves the highest success rate (72\%), followed by Claude (70\%) and Perplexity (66\%), with Copilot at 60\%. These percentages represent the proportion of correct solutions generated by each LLM, including those needing minor code edits.

\textbf{Hypothesis Testing}: To assess each model's success rate, we conducted a one-tailed binomial test with baseline thresholds of 50\%, 60\%, and 70\%, determining if each LLM’s success rate significantly exceeded these benchmarks. This non-parametric test, suitable for binary outcomes (correct/incorrect), provides insight into each LLM’s performance relative to random chance \cite{wohlin2012experimentation}. Additionally, we evaluated whether there was a significant difference in success rates between the LLMs by applying the Friedman test, followed by pairwise Wilcoxon tests where a significant difference was detected.

\begin{table}[H]
\centering
\caption{RQ1: Success rate results of LLMs at different baselines}
\label{tab:llm_success_rate}
\scriptsize 
\begin{tabular}{|c|c|c|c|c|}
\hline
\textbf{Baseline} & \textbf{LLM} & \textbf{Success Rate (\%)} & \textbf{p-value} & \textbf{Conclusion} \\ \hline
\multirow{4}{*}{50\%} 
    & Copilot     & 60\% & 0.0284 & Significant       \\ \cline{2-5} 
    & ChatGPT     & 72\% & 0.0000 & Significant       \\ \cline{2-5} 
    & Perplexity  & 66\% & 0.0009 & Significant       \\ \cline{2-5} 
    & Claude      & 70\% & 0.0000 & Significant       \\ \hline

\multirow{4}{*}{60\%} 
    & Copilot     & 60\% & 0.5433 & Not Significant   \\ \cline{2-5} 
    & ChatGPT     & 72\% & 0.0084 & Significant       \\ \cline{2-5} 
    & Perplexity  & 66\% & 0.1303 & Not Significant   \\ \cline{2-5} 
    & Claude      & 70\% & 0.0248 & Significant       \\ \hline

\multirow{4}{*}{70\%} 
    & Copilot     & 60\% & 0.9875 & Not Significant   \\ \cline{2-5} 
    & ChatGPT     & 72\% & 0.3768 & Not Significant   \\ \cline{2-5} 
    & Perplexity  & 66\% & 0.8371 & Not Significant   \\ \cline{2-5} 
    & Claude      & 70\% & 0.5491 & Not Significant   \\ \hline
\end{tabular}
\end{table}

As Table \ref{tab:llm_success_rate} shows, all LLMs perform significantly above the 50\% threshold, confirming baseline effectiveness in solving coding tasks. At the 60\% baseline, only ChatGPT and Claude reach statistical significance, suggesting enhanced reliability for typical tasks. No LLM achieves significance at the 70\% baseline, indicating limitations in sustaining very high success rates across diverse challenges.

\begin{figure}[H]
    \centering
    \includegraphics[width=0.75\linewidth]{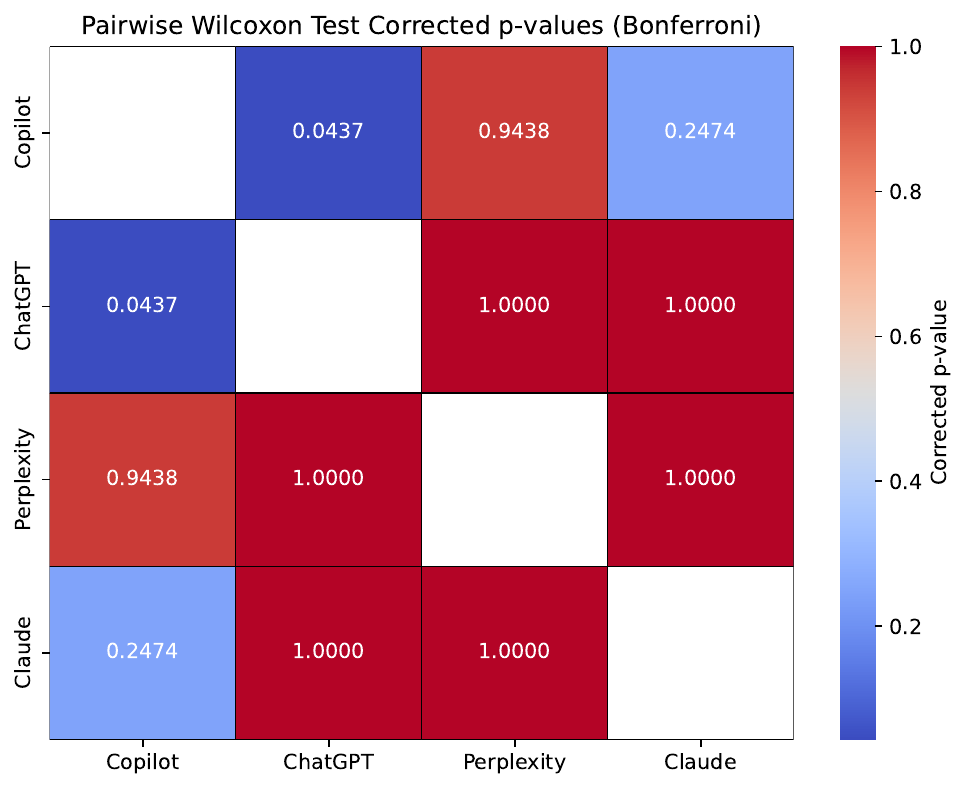}
    \caption{RQ1 - Pairwise Comparison of Success Rates.}
    \label{fig:rq1heatmap}
\end{figure}

To explore differences between LLMs, we applied the Friedman test, which detected significant variation in success rates across models (p = 0.0384). We followed up with post-hoc Wilcoxon pairwise comparisons, identifying a statistically significant difference between ChatGPT and Copilot, with ChatGPT achieving a significantly higher success rate (corrected p-value: 0.0437), as depicted in the heatmap of Figure \ref{fig:rq1heatmap}. No other significant differences were observed among models. 
%provides a heatmap visualization of the pairwise comparison results.

% \subsubsection{Discussion}
% The hypothesis tests for RQ1 support the following conclusions:
% \begin{itemize}
%     \item \textbf{50\% Baseline:} All LLMs demonstrate success rates significantly above 50\%, confirming they perform better than random chance.
%     \item \textbf{60\% Baseline:} Only ChatGPT and Claude surpass the 60\% baseline, underscoring these models’ relative reliability in data science coding tasks.
%     \item \textbf{70\% Baseline:} No LLM achieves significance at the 70\% baseline, highlighting a potential limitation in sustaining very high success rates across varied problems.
%     \item \textbf{Friedman Test and Wilcoxon Post-hoc Test:} Significant differences between LLMs were observed, with ChatGPT showing a significantly higher success rate compared to Copilot.
% \end{itemize}

% \subsubsection{Discussion}
Based on these tests, we conclude:
\begin{quote}
\textit{
For hypotheses \( H0_1 \) and \( H0_{1a} \):
\begin{itemize}
    \item At the \textbf{50\% baseline}, all LLMs exhibit success rates significantly above 50\%, supporting the conclusion that each model performs better than random chance in solving data science coding problems.
    \item At the \textbf{60\% baseline}, only ChatGPT and Claude show success rates significantly above this level, indicating that these two models exhibit greater reliability across general coding tasks.
    \item At the \textbf{70\% baseline}, no LLM meets statistical significance, suggesting a possible limitation in achieving consistently high success rates across diverse coding challenges.
    \item \textbf{Friedman Test and Wilcoxon Post-hoc Test:} Significant differences were found between models, with ChatGPT achieving a success rate significantly higher than that of Copilot.
\end{itemize}
}
\end{quote}

In summary, RQ1 indicates that ChatGPT and Claude exhibit the most consistent performance, particularly ChatGPT, which leads in relative success. These findings suggest that ChatGPT and Claude may be preferable for tasks demanding higher success rates, while highlighting the difficulty for LLMs in consistently achieving a 70\% success rate across diverse challenges.

\subsection{RQ2: Does the difficulty level of coding problems (easy, medium, hard) influence the success rate of the different LLMs?}

\begin{figure}[H]
    \centering
    \includegraphics[width=0.95\linewidth]{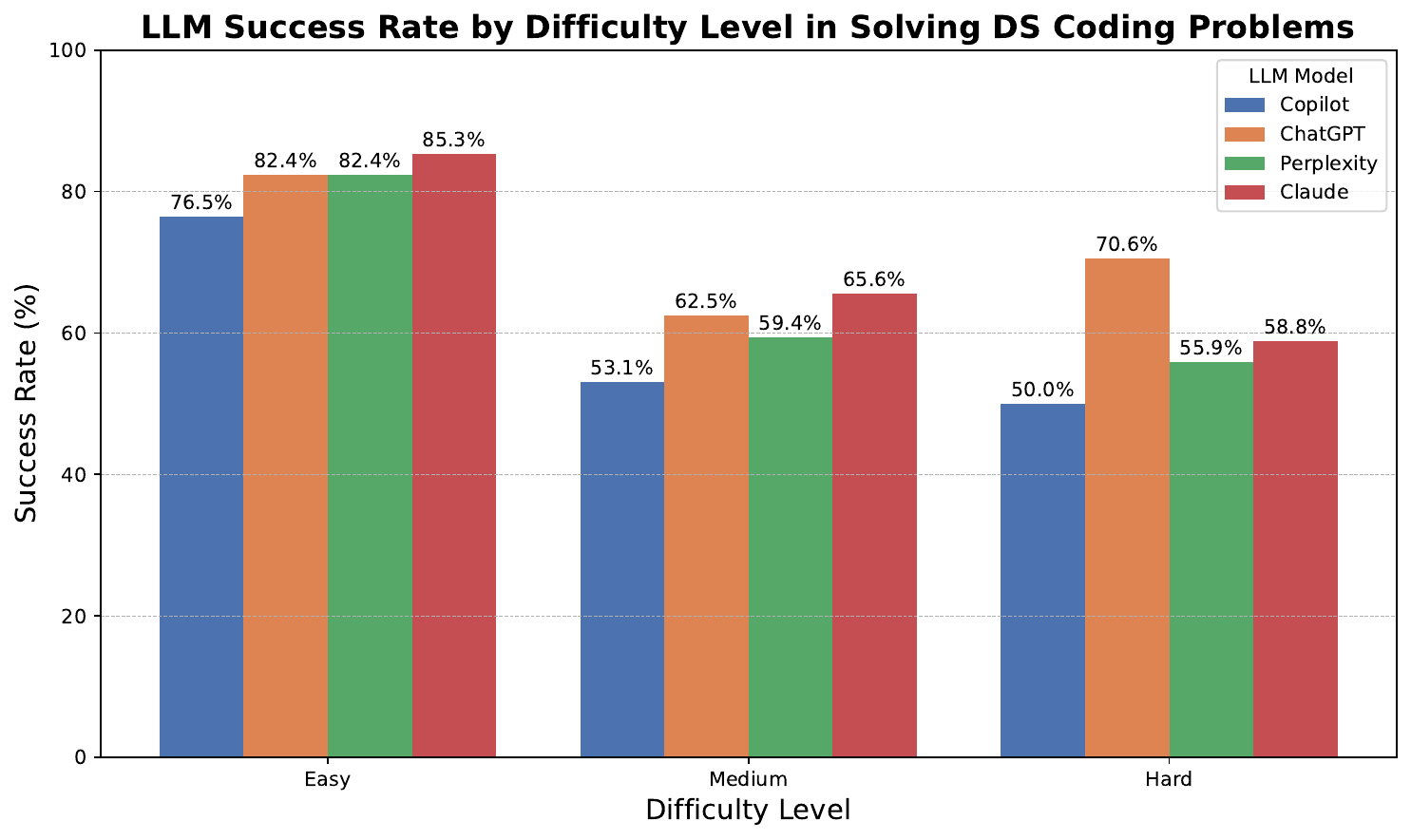}
    \caption{RQ2: Effect of difficulty level on success rate.}
    \label{fig:rq2}
\end{figure}

As shown in Figure \ref{fig:rq2}, the success rates of each LLM vary across different difficulty levels. Claude achieves the highest success rate on easy and medium problems, while ChatGPT excels on hard problems, suggesting its robustness with advanced challenges. Copilot consistently shows the lowest success rate across all difficulty levels, indicating a potential limitation in handling more complex tasks.

\textbf{Hypothesis Testing}: Chi-Square tests were performed to evaluate the effect of difficulty level on each LLM's success rate. Results show that difficulty level significantly impacts the success rates of Perplexity and Claude (\( p < 0.05 \)), suggesting that their performance fluctuates with problem complexity. In contrast, Copilot and ChatGPT demonstrate consistent success rates across all difficulty levels, indicated by non-significant results.

Based on these tests, we conclude:
\begin{quote}
\textit{For the hypothesis H0\_2, we reject it for Perplexity and Claude, indicating that difficulty level significantly affects their success rates. For Copilot and ChatGPT, we fail to reject H0\_2, suggesting consistent performance across varying difficulty levels.}
\end{quote}

To further explore comparative performance, we conducted the Friedman test across all models at each difficulty level. Although the overall test did not show significant differences in success rates across LLMs for each level, pairwise Wilcoxon tests highlighted a significant difference between ChatGPT and Copilot for hard problems (p = 0.0196), indicating ChatGPT's superior performance on more challenging tasks.

% \subsubsection{Discussion}
% The findings in RQ2 reveal that both Claude and ChatGPT maintain relatively high success rates across easy and medium problems, with ChatGPT displaying a distinct advantage on harder problems. Perplexity and Claude showed significant variability with problem difficulty, while ChatGPT and Copilot exhibited consistency, unaffected by the level of difficulty. These insights suggest that ChatGPT may offer greater adaptability to task complexity, particularly for challenging data science problems. However, Copilot's lower success rate across all levels highlights potential limitations in effectively scaling to more complex coding tasks.

% The Chi-Square test is used here because it is appropriate for categorical data, such as the counts of correct and incorrect solutions across difficulty levels. This non-parametric test evaluates whether there is a significant association between difficulty level and accuracy without requiring assumptions about data distribution.

\subsection{RQ3: Does the type of data science task (Analytical, Algorithm, Visualization) influence the success rate of the different LLMs?}
\begin{figure}[!ht]
    \centering
    \includegraphics[width=0.95\linewidth]{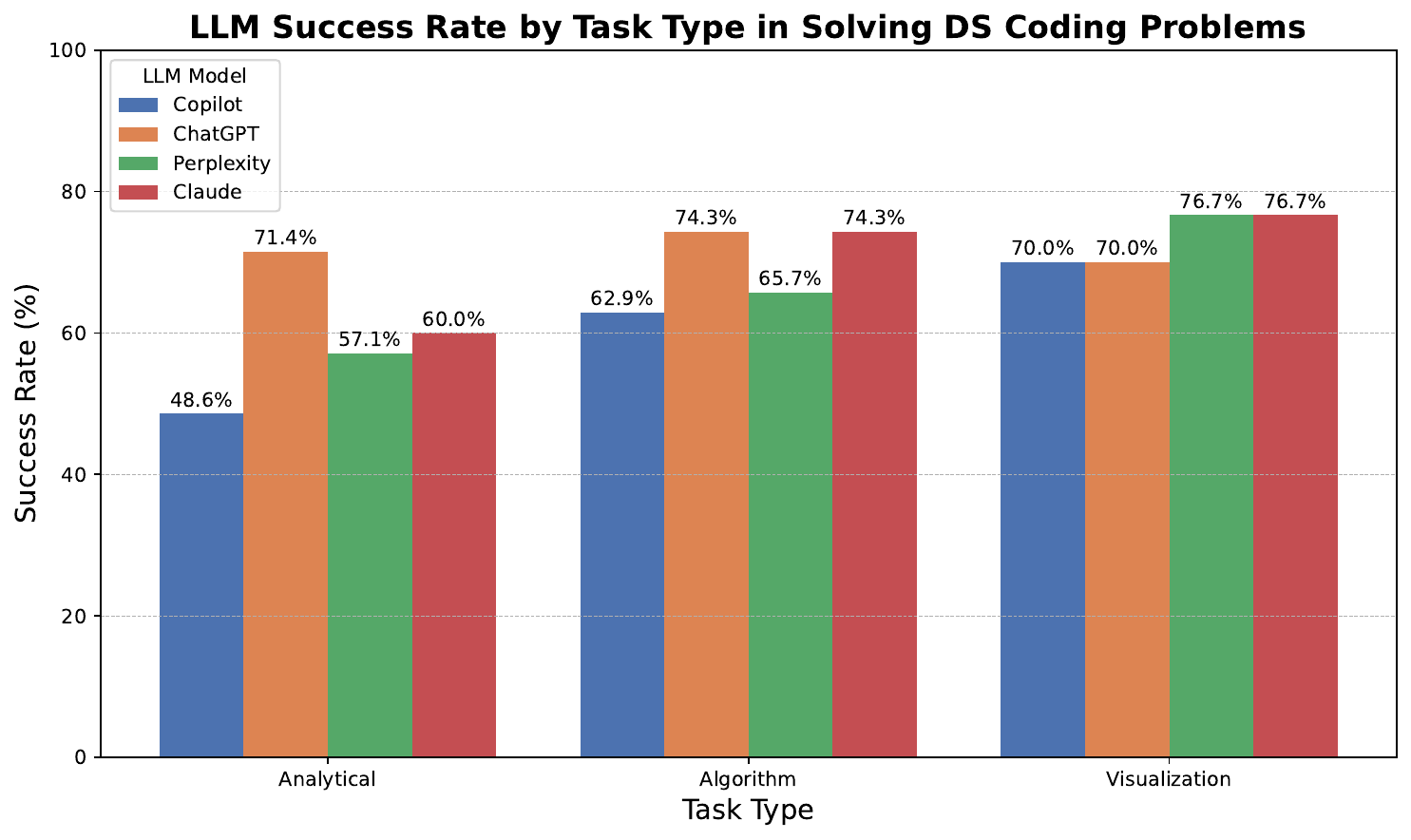}
    \caption{RQ3: Effect of task type on success rate.}
    \label{fig:rq3}
\end{figure}

Figure \ref{fig:rq3} illustrates the success rate of each LLM across different task types. ChatGPT demonstrates the highest success rate in analytical and algorithm tasks, while Perplexity and Claude achieve similar levels in visualization tasks. Although ChatGPT performs particularly well in analytical and algorithm tasks, statistical tests reveal no significant overall success rate differences among the models across task types, except between ChatGPT and Copilot.

\begin{table}[H]
\centering
\caption{RQ3: Chi-Square test results for task type across LLMs}
\label{tab:chi_square_task_type}
\scriptsize
\begin{tabular}{|c|c|c|c|c|c|}
\hline
\textbf{LLM} & \textbf{Algorithm} & \textbf{Analytical} & \textbf{Visuali.} & \textbf{p-value} & \textbf{Conclusion} \\ \hline
Copilot     & 22/35       & 17/35       & 21/30       & 0.1946   & Not Sig. \\ \hline
ChatGPT     & 26/35       & 25/35       & 21/30       & 0.9250   & Not Sig. \\ \hline
Perplexity  & 23/35       & 20/35       & 23/30       & 0.2534   & Not Sig. \\ \hline
Claude      & 26/35       & 21/35       & 23/30       & 0.2715   & Not Sig. \\ \hline
\end{tabular}
\end{table}

The Chi-Square test results in Table \ref{tab:chi_square_task_type} show that task type does not significantly impact the success rate for any LLM, with all p-values exceeding the 0.05 threshold. This finding suggests that each model’s performance remains relatively stable across analytical, algorithmic, and visualization tasks.

%To further explore potential differences, we conducted pairwise comparisons among the LLMs. While most comparisons showed no statistically significant differences, a notable exception emerged between ChatGPT and Copilot for analytical and algorithm tasks. This result allows us to affirm that ChatGPT significantly outperforms Copilot specifically for these types of tasks.

\begin{quote} \textit{For hypothesis H0\_3, we fail to reject it for all models, indicating that task type does not significantly impact success rate overall. However, post-hoc comparisons reveal that ChatGPT performs significantly better than Copilot in analytical and algorithm tasks.} \end{quote}

\subsection{RQ4: For Analytical questions, do the LLMs differ in the efficiency (running time) of the code they generate?}

\begin{figure}[H]
    \centering
    \includegraphics[width=1.0\linewidth]{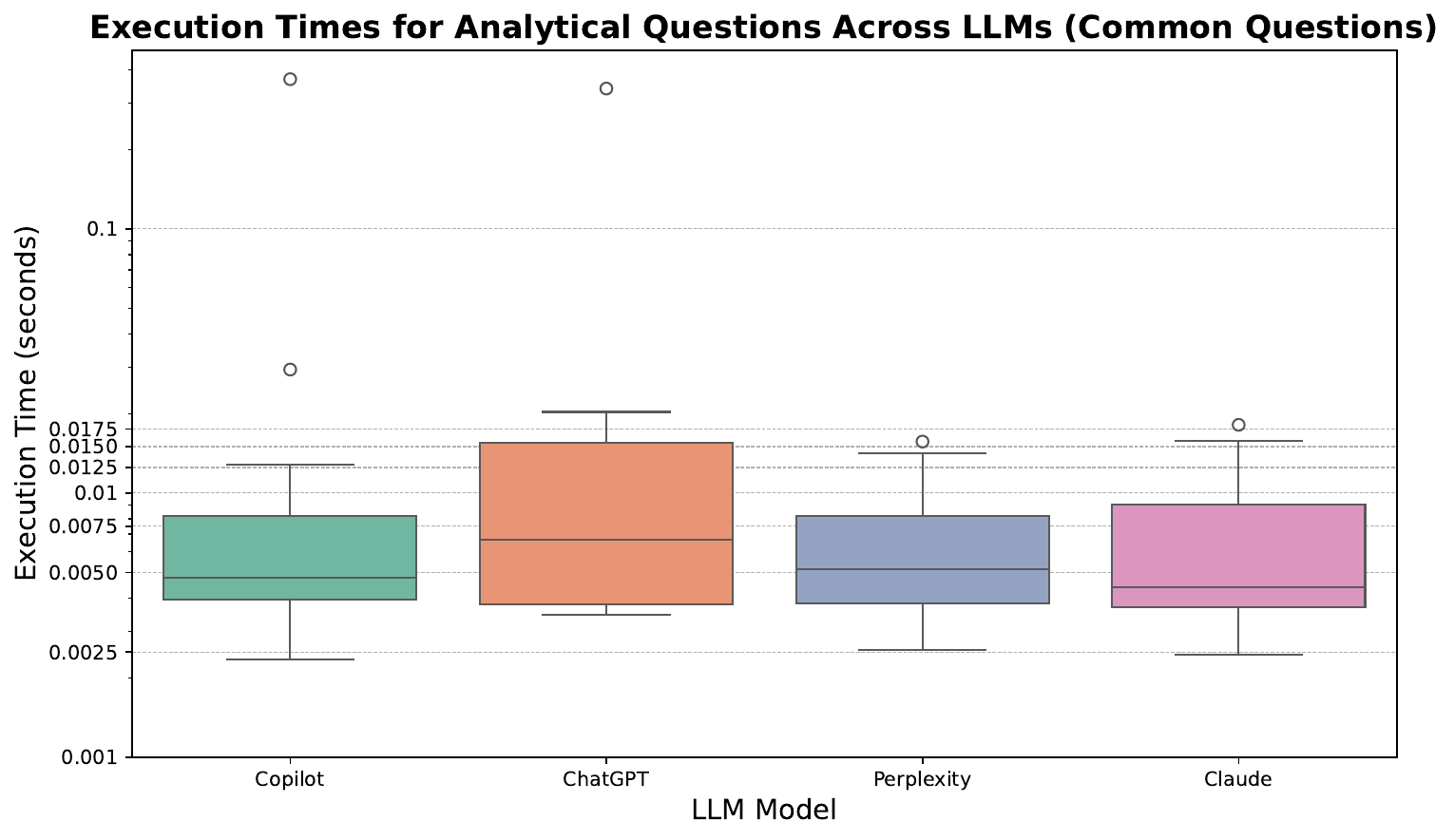}
    \caption{RQ4: Execution times of LLMs - Box Plot.}
    \label{fig:rq4}
\end{figure}

\begin{figure}[H]
    \centering
    \includegraphics[width=0.9\linewidth]{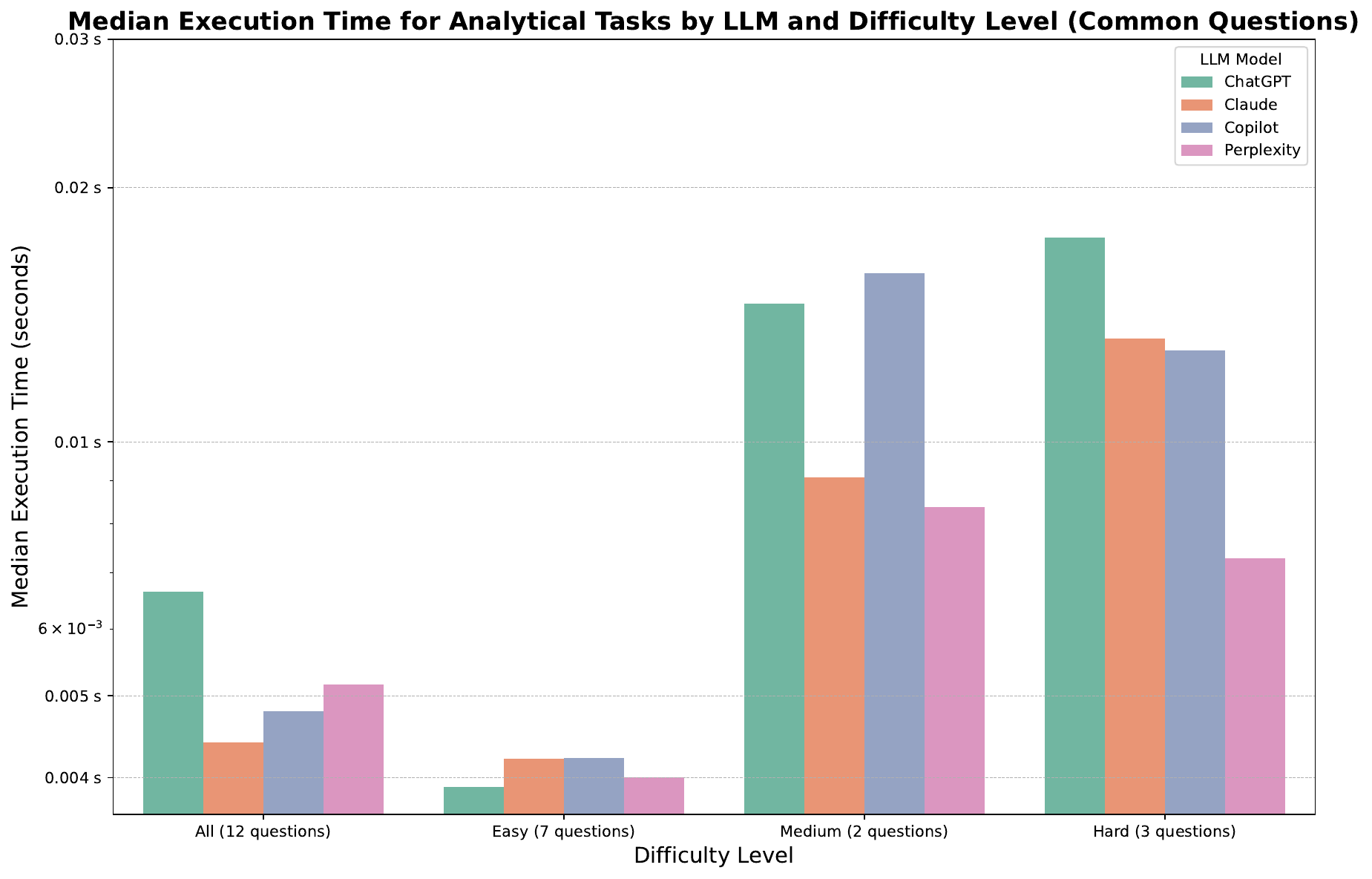}
    \caption{RQ4: Median execution time by difficulty level.}
    \label{fig:rq4-2}
\end{figure}

For a fair comparison, Figures \ref{fig:rq4} and \ref{fig:rq4-2} include only the results from problems successfully solved by all LLMs, as the platform does not compute execution times for solutions that did not work. Accordingly, Claude has the lowest median execution time, indicating its solutions generally execute faster than the other models' solutions, followed by Copilot and Perplexity. ChatGPT has the highest median execution time, suggesting that on average, it takes longer to execute analytical tasks than the other models. ChatGPT displays the largest interquartile range (IQR), indicating significant variability, whereas Copilot, Perplexity, and Claude have narrower IQRs, suggesting more consistent execution times. This analysis suggests Claude is generally faster and more consistent for analytical tasks, while ChatGPT may offer less predictability in execution time.

%Despite ChatGPT perform
%This analysis suggests that while Copilot may be faster on simpler tasks, it struggles with consistency on harder ones, making it less reliable for complex analytical tasks. Claude's and Perplexity's performances are more consistent across difficulty levels. ChatGPT, despite its robust performance in success rates, exhibits slower and less predictable execution times. %, which may affect efficiency in efficiency-sensitive workflows.

%As illustrated in Figure \ref{fig:rq4}, Copilot has the lowest median execution time, followed by Claude, Perplexity, and then ChatGPT, which has the highest median. Copilot and Claude demonstrate more consistency with less variability, while ChatGPT and Perplexity exhibit greater variability and occasional high execution times. Both Copilot and ChatGPT are prone to outliers, affecting reliability in specific tasks. On average, Copilot performs the quickest, while ChatGPT tends to be slower. ChatGPT displays the largest interquartile range (IQR), indicating significant variability, whereas Claude and Copilot have narrower IQRs, suggesting more consistent execution times. Copilot's distribution concentrates heavily near the lower end with a sharp peak, while ChatGPT and Perplexity feature wider distributions, consistent with their higher variability. Claude's distribution is symmetrical and concentrated around its median. Additionally, ChatGPT and Perplexity have longer tails, indicating a higher frequency of elevated execution times compared to Claude and Copilot.

\textbf{Hypothesis Testing}: To assess whether these observed differences are statistically significant, we conducted a Kruskal-Wallis test, as the Kruskal-Wallis test is a non-parametric method suitable for comparing the distributions of independent groups, particularly their central tendencies, when data is not normally distributed. %Given that the Shapiro-Wilk test confirmed that the execution times for each LLM do not follow a normal distribution, the Kruskal-Wallis test provides a reliable method for analyzing central tendencies (medians) without the need for normality.

% The hypotheses tested were:
% \begin{itemize}
%     \item Null Hypothesis (H0\_4): The population medians of the execution times across the LLMs for Analytical questions are equal.
%     \item Alternative Hypothesis (H1\_4): At least one LLM has a different population median execution time for Analytical questions compared to others.
% \end{itemize}

The test, conducted using the \texttt{scipy.stats} library, resulted in a Kruskal-Wallis statistic of 0.6947 and a p-value of 0.8744. With a p-value exceeding the significance level of 0.05, we fail to reject the null hypothesis. This suggests that there are no statistically significant differences in the median execution times across the LLMs for Analytical questions.

\begin{quote}
\textit{For RQ4, we fail to reject H0\_4, indicating that the LLMs do not differ significantly in the efficiency (running time) of the code they generate for Analytical questions.}
\end{quote}

\subsection{RQ5: For visualization tasks, do the LLMs differ in the quality (similarity) of the visual outputs they produce compared to expected results?}

\begin{figure}[!ht]
    \centering
    \includegraphics[width=1.0\linewidth]{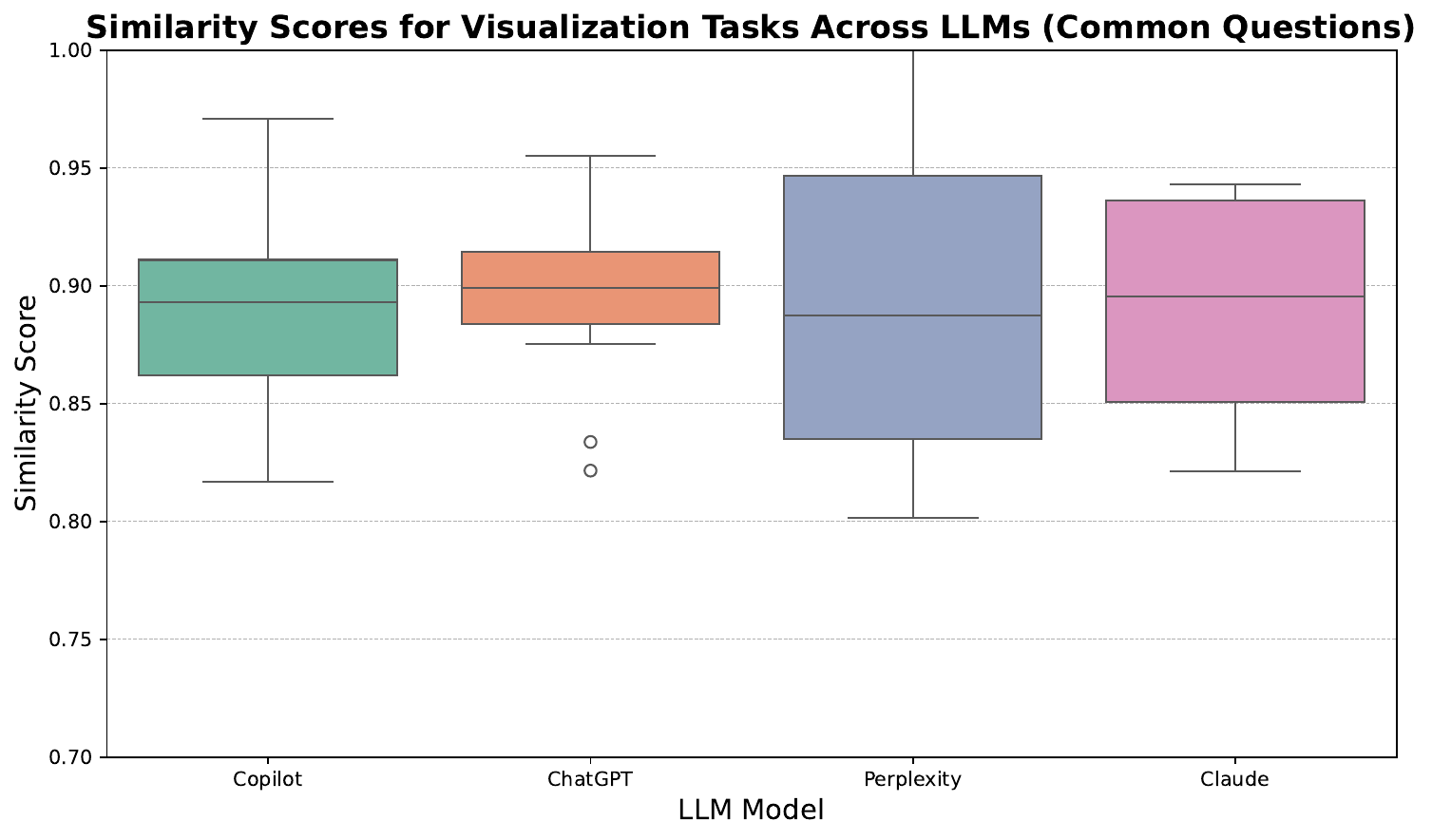}
    \caption{RQ5: Similarity Scores - Box Plot.}
    \label{fig:rq5}
\end{figure}

\begin{figure}[!ht]
    \centering
    \includegraphics[width=1.0\linewidth]{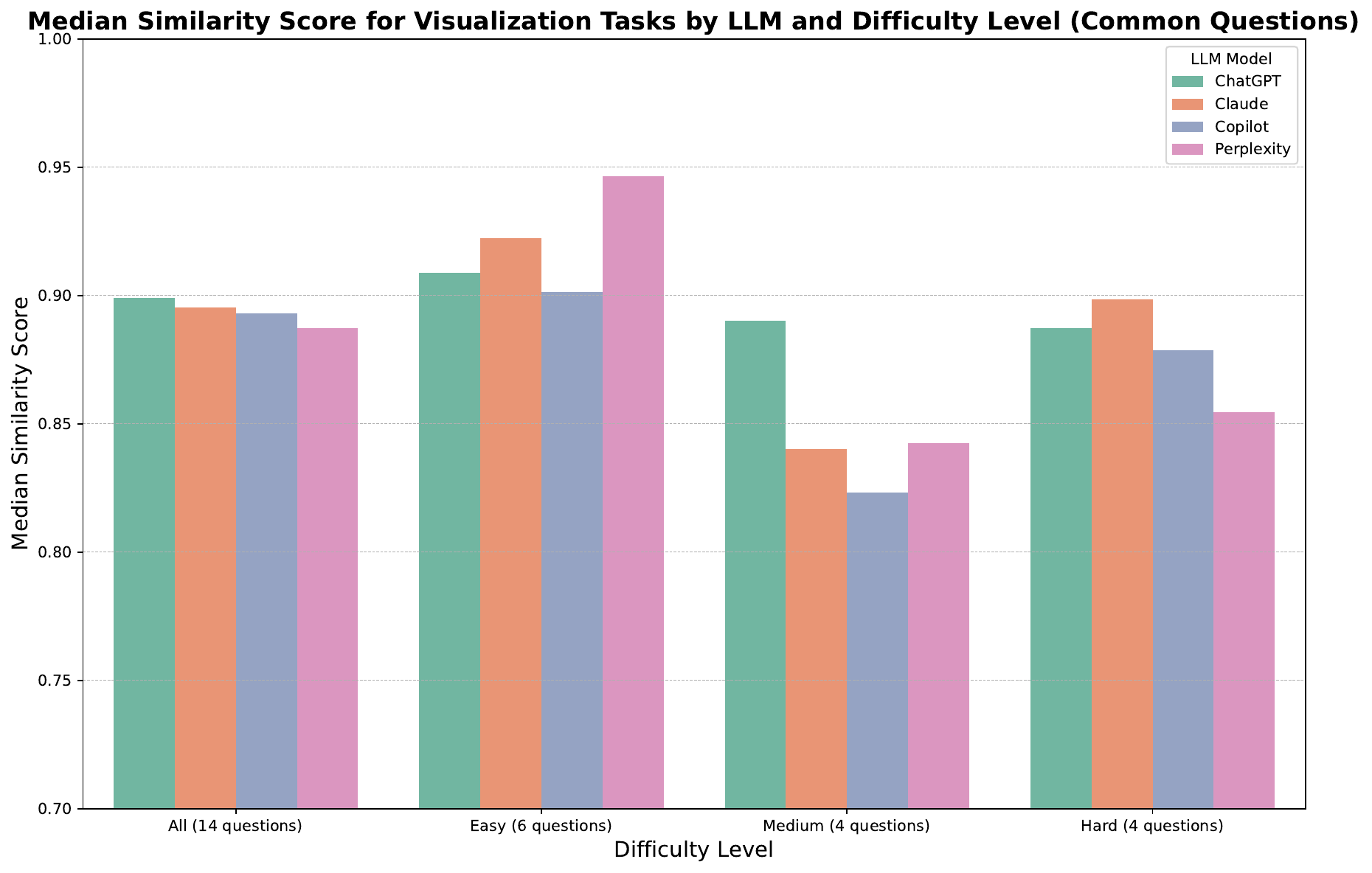}
    \caption{RQ5: Median similarity scores by difficulty level.}
    \label{fig:rq5-2}
\end{figure}

% \begin{figure}[!ht]
%     \centering
%     \includegraphics[width=1.0\linewidth]{figures/results/RQ5-0.pdf}
%     \caption{RQ5: Similarity Scores distribution.}
%     \label{fig:rq5-0}
% \end{figure}

As depicted in Figures \ref{fig:rq5} and \ref{fig:rq5-2}, ChatGPT achieves the highest median similarity score among the commonly solved problems, indicating that its outputs are closest to the expected results. Additionally, ChatGPT displays the narrowest interquartile range (IQR), highlighting its consistency. These findings suggest that ChatGPT delivers more reliable quality in generating visual outputs that closely match the expected results.

\textbf{Hypothesis Testing}: To statistically analyze differences in similarity scores among the LLMs, we conducted a Kruskal-Wallis test. 
%, as similarity scores are not assumed to be normally distributed across all LLMs. 
%This non-parametric test compares the medians across groups, which is particularly useful given the observed variability in ChatGPT's performance.

\textbf{Kruskal-Wallis Test Results}:  
\begin{itemize}
    \item Kruskal-Wallis Statistic: 0.8287  
    \item p-value: 0.8426  
    \item \textbf{Conclusion}: The p-value above 0.05 suggests no statistically significant differences in similarity scores between the LLMs. This indicates that while there are observed differences in mean similarity scores and variability (with ChatGPT achieving the highest mean and most consistent performance), these differences are not statistically significant across LLMs at the 5\% significance level.
\end{itemize}

Based on these results, we conclude:
\begin{quote}
\textit{For hypothesis H0\_5, we fail to reject the null hypothesis, indicating that there is no significant difference in the similarity quality of generated visualization outputs among the LLMs.}
\end{quote}

\subsection{Threats to Validity}

As usual in empirical studies, our study acknowledges several threats that may impact the interpretation and generalization of the results. 

\subsubsection{Internal Validity}

A key concern is the undisclosed nature of the LLMs' training data. Without access to this information, we cannot confirm whether the generated solutions are novel or based on memorized content. Even though we selected new problems from Stratascratch, similar or identical problems might exist in the models' training data, potentially inflating their apparent effectiveness.

Prompt design is another factor influencing outcomes. As noted by White et al. \cite{white2023chatgpt}, the formulation of prompts can significantly affect LLM outputs. While we endeavored to use consistent prompts derived from original problem descriptions, variations could lead to different results.

To address potential subjectivity in converting problems to prompts, we developed standardized prompt templates for each task type. These templates ensured that all AI assistants received clear, consistent instructions, allowing for a fair comparison of performance.

\subsubsection{External Validity}

The generalizability of our findings is limited by the scope of problems used. Our study focused on 100 Python coding problems from a single platform, which may not represent the full spectrum of data science tasks. To enhance external validity, future research should incorporate a wider range of problems from multiple sources.

\subsubsection{Construct Validity}

We did not formally assess the expertise of the researchers conducting the experiment, which could introduce subjectivity, particularly in interpreting and evaluating the AI-generated code. Although guidelines were established for acceptable code modifications—allowing only minor edits to resolve execution issues—differences in coding proficiency among researchers could influence the assessment.

\subsubsection{Conclusion Validity}
These threats may affect the validity of our conclusions. While our study offers insights into the capabilities and limitations of LLMs in data science code generation, the results should be interpreted with caution. Further research addressing these limitations is necessary to strengthen the confidence in the findings.

% \section{Conclusion} \label{sec:conclusion}

% This paper presented a pioneering controlled experiment evaluating the capabilities of four prominent LLM-based AI assistants—Microsoft Copilot (GPT-4 Turbo), ChatGPT (o1-preview), Claude (3.5 Sonnet), and Perplexity Labs (Llama-3.1-70b-instruct)—in solving data science-specific coding tasks. By systematically assessing these models on a curated set of data science problems from the Stratacratch platform, we aimed to understand their effectiveness across different task types and difficulty levels.

% \textcolor{red}{ In summary, the result from the data analysis...
% The results highlight both the potential and limitations of LLMs in addressing data science challenges. While }

\section{Discussion} \label{sec:dicussion}
Through a series of hypothesis tests, we investigated each model's effectiveness across different problem types and difficulty levels. The findings underscore both the strengths and limitations of these LLMs in addressing data science challenges, providing insights into which models may be most suitable for specific scenarios in data science workflows. The results highlight that:
\begin{itemize}
    \item \textbf{Success Rate:} Empirical evidence from our tests indicates that each LLM exceeded the 50\% baseline success rate, confirming effectiveness beyond random chance. At the 60\% baseline, only ChatGPT and Claude achieved significantly higher success rates, reinforcing their reliability in general coding contexts. However, none of the models reached the 70\% threshold, suggesting limitations in consistently achieving high accuracy across diverse data science task types. ChatGPT achieved the highest overall success rate and performed consistently well on harder questions, with descriptive analysis suggesting strong outcomes in analytical and algorithmic tasks, reflecting its robustness in complex data science scenarios. Claude also demonstrated solid performance, particularly on easier and medium-difficulty tasks, as well as in visualization tasks, indicating versatility across various problem types. Perplexity and Copilot, while showing lower success rates on more complex tasks, displayed consistent performance on simpler tasks, highlighting their potential for straightforward data science workflows. 
    
    \item \textbf{Efficiency (Execution Time):} For analytical tasks, the Kruskal-Wallis test on execution times revealed no statistically significant differences among the models, suggesting that efficiency, in terms of runtime, is relatively comparable across these LLMs. This finding implies that while execution time may vary, it may not be a decisive factor in model selection for tasks where accuracy and complexity are primary concerns.
    Despite the lack of empirical significance, the median execution times indicate some practical trends: Claude had the lowest median execution time, suggesting it generally runs faster than the other models, followed by Copilot and Perplexity. ChatGPT had the highest median execution time, indicating slower performance on average. 
    %Additionally, ChatGPT displayed the largest interquartile range (IQR), showing significant variability, while Copilot, Perplexity, and Claude had narrower IQRs, implying more consistent execution times. This suggests that Copilot is generally faster and more consistent for analytical tasks, whereas ChatGPT may offer less predictability in execution speed.

    \item \textbf{Quality of Output (Image Similarity for Visualization):} In visualization tasks, where models were evaluated based on similarity scores to expected outputs, ChatGPT achieved the highest median similarity score. However, statistical tests indicated no significant differences between the models. %, showing that all LLMs possess similar baseline capabilities for generating visual outputs, though Claude may offer slightly better consistency in output quality.
    \item \textbf{Consistency Across Difficulty Levels and Task Types:} Empirical analysis reveals that ChatGPT maintains consistent performance regardless of task complexity, providing reliable success rates across both simple and complex tasks. In contrast, Perplexity's and Claude's success rates were significantly influenced by task difficulty, with better outcomes on less complex tasks. Copilot also demonstrated consistency across difficulty levels, though with generally lower success rates than ChatGPT. Additionally, our tests indicate that task type (analytical, algorithmic, visualization) does not significantly impact success rates for any model, suggesting stable performance across different data science task types. This consistency makes ChatGPT a dependable choice when task complexity is uncertain.
\end{itemize}

\section{Conclusion} \label{sec:conclusion}

% This paper presented a controlled experiment evaluating the effectiveness of four prominent LLM-based AI assistants—Microsoft Copilot (GPT-4 Turbo), ChatGPT (o1-preview), Claude (3.5 Sonnet), and Perplexity Labs (Llama-3.1-70b-instruct)—in solving data science coding tasks. Effectiveness in this context was defined as the models' ability to achieve high success rates (correct solutions), demonstrate efficient execution (speed), produce quality visual outputs (similarity), and maintain consistent performance across varying levels of difficulty and task types. 

This study presents a controlled experiment evaluating the effectiveness of four prominent LLM-based AI assistants—Microsoft Copilot (GPT-4 Turbo), ChatGPT (o1-preview), Claude (3.5 Sonnet), and Perplexity Labs (Llama-3.1-70b-instruct)—in data science coding tasks. Effectiveness was measured by each model's success rate, execution efficiency, visual output quality, and consistency across difficulty levels and task types.

With success rates exceeding 50\% for all models, this research provides valuable insights into LLM performance in data science. At the 60\% baseline, only ChatGPT and Claude achieved significantly higher success rates, highlighting their reliability in general coding tasks. However, our findings indicate that only ChatGPT consistently maintains performance across different difficulty levels, whereas Claude's success rate is significantly affected by task difficulty, suggesting its performance may vary with more complex tasks.

No evidence suggests that task type affects LLM success rates, though ChatGPT (o1-preview) significantly outperforms Copilot (GPT-4o) for analytical and algorithm tasks. This nuanced understanding of each model's strengths enables more strategic LLM selection tailored to specific needs. Additionally, this study underscores the value of rigorous hypothesis testing in AI evaluation, setting a template for assessing models beyond basic accuracy metrics.

\section{Future Work} \label{sec:threats}
Our study opens several avenues for future research to enhance the application of LLMs in data science code generation.

\subsection{Exploring Complex and Real-World Data Science Tasks}
Evaluating LLMs on sophisticated, real-world data science tasks—such as implementing machine learning models with libraries like Scikit-learn or TensorFlow, handling large datasets, and working with unstructured data—could provide deeper insights into their capabilities and limitations. For instance, Nascimento et al. \cite{nascimento2023gptloop} demonstrated the use of an LLM to replace a learning algorithm that involved neural networks optimized through genetic algorithms. While their experiment was preliminary, it highlighted the potential of LLMs to automate complex coding solutions, suggesting that these models could extend beyond basic scripting to more advanced tasks. Exploring tasks like multivariate analysis, time series forecasting, and dynamic optimization could further test LLM proficiency. Testing in practical settings uncovers challenges that controlled experiments may not fully capture.

\subsection{Expanding Model Diversity and Dataset Coverage}

We could extend this analysis by integrating additional LLMs and incorporating data science-specific coding challenges from various platforms, such as LeetCode, with tasks like data manipulation, cleaning, and SQL queries. To capture a broader range of data science skills, the dataset could also include non-coding tasks, such as interpretation and analysis questions, as provided by Stratascratch.

Additionally, we could integrate recently released questions in Stratascratch that use Polars DataFrame \cite{polars2023} for data manipulation-a high-performance library designed for efficient data handling in Python.
%This addition would test the models' adaptability to advanced libraries tailored for optimized data processing workflows. 
We could further expand the dataset by incorporating problems from DS-1000 \cite{lai2023ds}, which includes a diverse selection of data science problems sourced from StackOverflow. Following recommendations by Lai et al. \cite{lai2023ds}, introducing customized variations of existing problems would help reduce model memorization, enhancing the rigor of the evaluation environment.

% \subsection{Exploring non-coding Data Science problems}
% \textcolor{red}{Stratascratch also contains non-coding problems}

\subsection{Expanding Evaluation Metrics}
Future work could expand LLM evaluation by integrating software engineering metrics like code complexity, maintainability, and readability. Code similarity analysis could assess alignment with industry standards, while qualitative reviews by data scientists would add valuable insights, particularly for visualization tasks where image similarity metrics may fall short.

\subsection{Investigating Prompt Engineering and Ensuring Reproducibility}
Prompt engineering significantly influences LLM outputs. Future research should examine how different prompt formulations affect code generation quality and consistency. Employing methodologies where LLMs simulate multiple users \cite{aher2023using} could shed light on the impact of varying professional experiences and prompt designs. Addressing the non-deterministic nature of LLMs by controlling parameters like temperature settings could improve reproducibility, leading to more consistent and reliable evaluations.

\subsection{Exploring Further Research Questions}
Even using the same dataset, many additional questions and hypotheses remain to be explored. For instance, while we assessed the impact of problem difficulty and type on LLM success rates, further analysis could focus on establishing baseline success rates for each problem type and difficulty level.  Given the general success baseline of 60\%, future research might explore optimal baseline thresholds specific to each type and level of task.
Beyond success rates, this dataset also allows for an in-depth exploration of efficiency (execution times) and similarity scores for each difficulty level, providing a more comprehensive view of model performance in diverse task complexity. Additionally, information on the number of attempts (up to three) and instances of minor edits provides data for assessing error types (syntax and logic errors), retry patterns, and the models' adaptability to user feedback. Our dataset also includes specific topics within each question type, allowing for a more granular analysis that could reveal topic-specific strengths or limitations of each LLM. 

% \textcolor{red}{"How much can we improve the performance of Codex by tuning
% the temperature and prompts per a specific problem?" - }
% % \cite{doderlein4496380piloting}

% \subsection{Building on Preliminary Studies and Integrating with Data Connectors}
% Building upon preliminary work, such as that by Nascimento et al. \cite{nascimento2023gpt}, with more rigorous experiments can validate the viability of LLMs in automating complex algorithmic solutions within data science. Integrating generative models with data connectors-as in frameworks like LangChain-could reveal how different LLMs interact with data and perform in diverse contexts, enhancing their practical applicability.

% \section*{Acknowledgment}
% xxxxx anonymous now

\bibliographystyle{IEEEtran}
\bibliography{references}

\end{document}